# PIPELINED PARALLEL FFT ARCHITECTURE

Mr. Tanaji U. Kamble [1], Mr. Patil B.G [2], Ms. Rakhee S. Bhojakar [3]

[1 2 3]Electronics Engineering Department

[1]Assistant Professor,
[2]Associate Professor,
[3]M.Tech Student.

[1] tanajikamble13@gmail.com, [2] patilbg@rediffmail.com, [3] rakhee1089@gmail.com.

**ABSTARCT:**
In this paper, an optimized efficient VLSI architecture of a pipeline Fast Fourier transform (FFT) processor capable of producing the reverse output order sequence is presented. Paper presents Radix-2 multipath delay architecture for FFT calculation. The implementation of FFT in hardware is very critical because for calculation of FFT number of butterfly operations i.e. number of multipliers requires due to which hardware gets increased means indirectly cost of hardware is automatically gets increased. Also multiplier operations are slow that's why it limits the speed of operation of architecture. The optimized VLSI implementation of FFT algorithm is presented in this paper. Here architecture is pipelined to optimize it and to increase the speed of operation. Also to increase the speed of operation 2 levels parallel processing is used.

## 1.  INTRODUCTION

FFT algorithm is very much needed in digital signal processing applications like filtering spectrum analysis etc. FFT plays a prominent role in  advanced digital communications system e.g. digital video broadcasting, orthogonal frequency division multiplexing (OFDM) systems etc. and it also allows the realization of the widely used wireless protocols IEEE 802.11n (WLAN) and the IEEE 802.16 (WiMax)[1],[2]. FFT is a very efficient way algorithm used to compute the Discrete Fourier Transform (DFT), now days it is very much used in most operations of digital signal processing and in communication. There are number of FFT algorithms existing with each algorithm having very much arithmetic and also arithmetic consists of complex multiplication, addition and subtraction. The Fast Fourier Transform (FFT) is an efficient and best way to for finding out the DFT of a finite sequence and its computational complexity is very much less than  that of direct evaluation of DFT[3]. FFT takes the advantage of calculating coefficients of the DFT iteratively that is why computational complexity automatically gets reduced. The fundamental principle of FFT algorithm is to decompose the DFT computation of length of sequence *N* into smaller and smaller DFT. It is based on dividing and breaking the FFT length into smaller FFT length and finally adding them to get complete FFT. It reduces the time complexity to evaluate a discrete Fourier transform so that is why the performance of DFT calculation increases in factor of 100 or more over general way of calculation of DFT according to standard definition of DFT. The main job of Discrete Fourier Transform (DFT) is converting a sequence of values into of different frequency components. It is very much needed in many fields but computing it in general way from the standard definition is too much slow and difficult to be practical. An FFT is a best way to compute the same result in a fast way: finding a DFT of *N* points in the in general direct way, according to definition, takes computational complexity of O ( $N^2$ ), whereas an FFT can find out the same result with computational complexity of *O(N log N)* which is very less than O( $N^2$ ) for higher value of *N*. There are generally two classes of FFT algorithms which are most famous and mostly utilized. First one is Decimation in time (DIT) algorithm and second one is Decimation in frequency (DIF) algorithm [3]. In this paper Decimation in Frequency (DIF) algorithm is used which is capable of producing reverse output order when input sequence is in normal order. Much research has been carried out on designing pipelined architectures for computation of FFT of signals. No. of algorithms have been invented to reduce the computational complexity, among them Cooley-Tukey radix-2 FFT is very popular. On the basic radix-2 FFT approach algorithms radix-4, split-radix and radix- 2^2 have been developed [4]. In the literature based on these algorithms there are some standard architectures, R2MDC (Radix-2multi-path delay commutator) is one of the best architecture way for pipelined implementation of radix-2 FFT [5][6][7].In this Paper (R2MDC) architecture for computation of FFT is presented. Several parallel architectures for FFT have been proposed in the literature, architectures are developed for a specific -point FFT. A formal method of developing these architectures from the algorithms is not well developed. Further, most of these require high hardware complexity and hardware architectures are not fully utilized. This is the era of high speed digital communications, and it require high throughput and low power designs to achieve this speed power requirement while keeping the hardware as far as possible to a minimum. Few pipelined architectures for real valued signals have been proposed based on the Brunn algorithm. However, these are not widely used. Several algorithms have been proposed for computation of RFFT. These approaches are based on removing redundancies of the FFT when input is real. These can be efficiently used in a digital signal processor compared to a specialized hardware implementation.

The paper structure is as follows: II Section discusses previous work done on FFT architecture implementation. Section III discusses the FFT algorithm implementation (Cooley-Tukey) and complex multiplication used inside the butterfly-processing element. Section III devoted for an architectural description of the FFT used module. Section IV shows the resulting implementation and finally a conclusion is given in section V.

## 2. FFT ALGORITHM

In this section, a brief overview of IFFT and FFT algorithms is provided to be effectively used in OFDM applications. The N-point discrete Fast Fourier Transform (DFT) is defined as:

$$X(k) = \sum_{n=0}^{N-1} x(n) \cdot W_N^{nk} \qquad (1)$$

Where $W_N^{nk} = e^{\frac{-j2\pi nk}{N}}$   $0 \leq k \leq N-1$

$X(k)$ = k-th no. harmonic, whereas

$x(n)$ = n-th no. input sample applied to FFT algorithm.

Calculation of DFT according to standard definition requires a computational complexity of $O(N^2)$. By using FFT algorithm, this complexity can be decreased up to $O(N.\log_r N)$.

### 2.1. The Cooley-Tukey FFT Algorithm

The Cooley-Tukey FFT is the most popular and universal among all FFT algorithms, because any factorization of input length N is possible with Cooley-Tuckey. The ways of Cooley-Tukey FFTs are those were thelength of transformation is a power of a radix r, in short $N = r^s$. We refer this algorithm as radix-r algorithms. Radix-r algorithm with basis r = 2 and r = 4 are most commonly used. If r = 2 and S stages, then according to following index mapping technique of the Cooley–Tukey algorithm gives:

$$n = 2^{s-1} n_1 + 2^{s-2} n_2 + 2^{s-3} n_3 + \cdots + 2 n_{s-1} + n_s \qquad (2)$$

$$k = 2^{s-1} k_1 + 2^{s-2} k_2 + 2^{s-3} k_3 + \cdots + 2 k_{s-1} + k_s \qquad (3)$$

And: $n_1, n_2, n_3, \ldots, n_{s-1}, n_s = 0,1$

$k_1, k_2, k_3, \ldots, k_{s-1}, k_s = 0,1$

This Cooley–Tukey Algorithm is based on the principle of divide and conquers approach in the frequency domain and that is why we refer it as decimation-in-frequency (DIF) FFT. Here the DFT formula is converted into two factor summations as shown below:

$$X(k) = \sum_{n=0}^{\frac{N}{2}-1} x(n) \cdot W_N^{nk} + \sum_{n=\frac{N}{2}-1}^{N} x(n) \cdot W_N^{nk}$$

$$= \sum_{n=0}^{\frac{N}{2}-1} x(n) \cdot W_N^{nk} + \sum_{n=\frac{N}{2}-1}^{N} x\left(n + \frac{N}{2}\right) \cdot W_N^{(n+\frac{N}{2})k}$$

$$= \sum_{n=0}^{\frac{N}{2}-1} x(n) \cdot W_N^{nk} + \sum_{n=\frac{N}{2}-1}^{N} x\left(n + \frac{N}{2}\right) \cdot W_N^{nk} \cdot W_N^{\frac{N}{2}k}$$

Here $W_N^{\frac{N}{2}k} = (-1)^k$ therefore

$$X(k) = \sum_{n=0}^{\frac{N}{2}-1} \left(x(n) + (-1)^k x\left(n + \frac{N}{2}\right)\right) W_N^{nk}$$

Now here X(k) can be decimated into even-and odd-indexed Frequency samples:

$$X(2k) = \sum_{n=0}^{\frac{N}{2}-1} \left(x(n) + x\left(n + \frac{N}{2}\right)\right) W_{\frac{N}{2}}^{nk} \qquad (4)$$

$$X(2k+1) = \sum_{n=0}^{\frac{N}{2}-1} \left(x(n) - x\left(n + \frac{N}{2}\right)\right) W_{\frac{N}{2}}^{nk} \qquad (5)$$

If this computational procedure of decimation of the N/2-point DFTs X(2k) and DFTs X(2K+1) repeated, we will require $\log_2 N$ stages, where each stage involves N/2 operation units (butterflies). While calculating the N point DFT using the DIF (decimation-in-frequency) FFT algorithm, as in the DIT (decimation-in-time) algorithm requires $(N/2).\log_2 N$ complex multiplication and $N.\log_2 N$ complex addition.

### 2.2. FFT module

The flow graph of complete DIF decomposition of basic 8-point DFT computation is represented in Fig. 1.It consist of basic operation a butterfly; it's 2 point DFT computation as shown in Fig. 2. 8 point DIF FFT algorithm Radix-2 butterfly processor consists of a complex adder and complex subtraction and complex multiplier [8][9][10]. The operation of complex multiplication between flowing data in FFT algorithm with the twiddle factor requires four times real multiplications and two times add/subtract operations

### 2.3. R2MDC Architecture

Main necessities of FFT processor in communication systems are high throughput, modularity and simplicity. The pipeline architecture can be used to fulfill these requirements. If input data stream is sequential in pipeline architecture the data doesn't match the requirements of FFT algorithm. In this process data should be stored in memory and then it can be accessed as per requirement using control logic as shown in the fig.3 R2MDC architecture i.e. Radix-2 Multi-path Delay Commutator architecture is a ultimate way to implement pipelined parallel FFT architecture. In this architecture we rearrange the data easily as per requirement of FFT/IFFT algorithm as shown in Fig.1[8][9].In this architecture data is divided into two parallel streams flowing in the forward direction and entering the butterfly scheduled by proper delays and maintaining proper distance between two streams. In fig.3proposed 8 point Radix-2 multipath delay commutator (R2MDC) is shown. In this architecture at each stage of the architecture half of data is delayed using memory (reg.) and half of data is processed. AT each stage 4, 2 and 1 delay required respectively as shown in the fig.3 For 8 point FFTR2MDCarchitecture total 10 delay elements (4+2+2+1+1=10) require. Also it requires 2 switches as shown in the fig.3

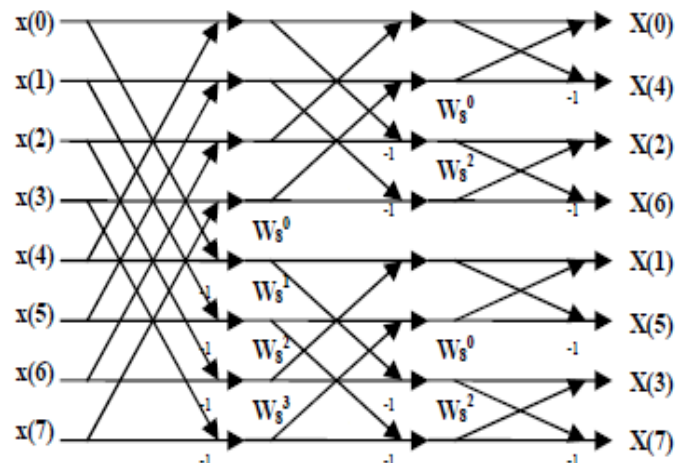

Fig.1. 8-point DIF (decimation-in-frequency) FFT algorithm

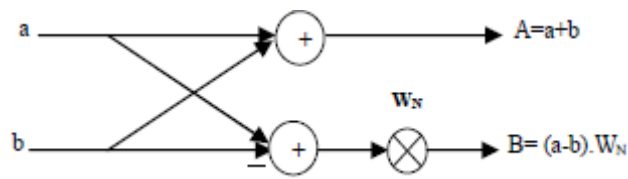

Fig.2. Butterfly operation

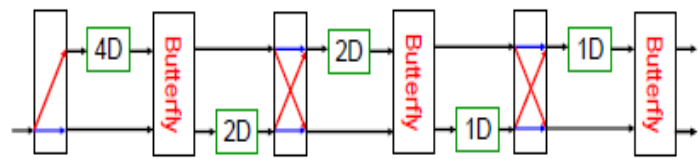

Fig.3. 8 point DIF R2MDC FFT architecture

# International Journal of Research in Advent Technology
Available Online at: http://www.ijrat.org

Main necessities of FFT processor in communication systems are high throughput, modularity and simplicity. The pipeline architecture can be used to fulfill these requirements. If input data stream is sequential in pipeline architecture the data doesn't match the requirements of FFT algorithm. In this process data should be stored in memory and then it can be accessed as per requirement using control logic as shown in the fig.3 R2MDC architecture i.e. Radix-2 Multi-path Delay Commutator architecture is a ultimate way to implement pipelined parallel FFT architecture. In this architecture we rearrange the data easily

**2.4.** *RTL of R2MDC Architecture*

In the fig. 4 RTL of 8 point DIF R2MDC FFT architecture is shown. This architecture is created using Xilinx software [10].Here sequential real data indicated by din_re and din_im is real part and imaginary part of input data stream respectively. FFT output is indicated by dout_re and dout_im, the real and imaginary part respectively. Here all the value for din_re are set to 2 and all the value for dout_im are set 2. The corresponding result is as shown in the fig.5

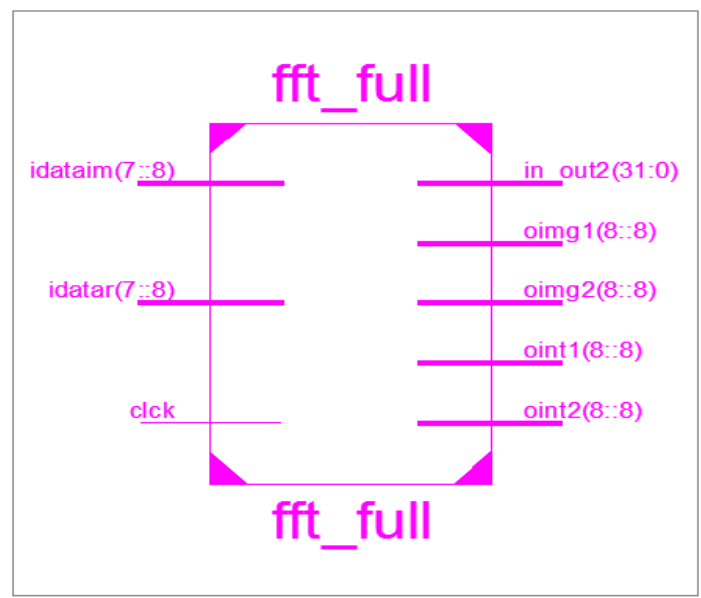

Fig.4. RTL of 8 point DIF R2MDC FFT architecture

**2.5.** *Simulation results*

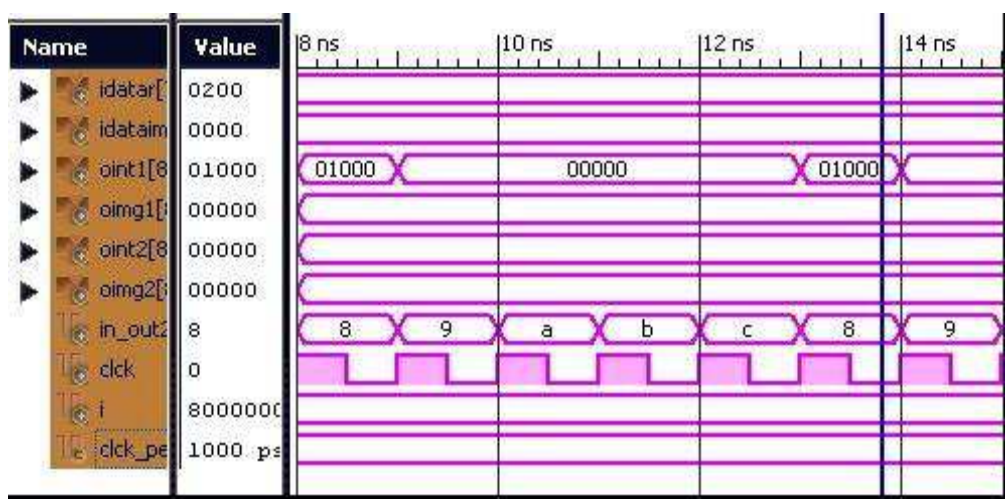

Fig.5. Simulation result of 8 point DIF R2MDC FFT architecture.

Table I: Comparison of the proposed architecture with Cooley-Tukey algorithm.

| Metrics | R2MDC architecture | Cooley-Tucky architecture |
|---|---|---|
| Total REAL time to Xst completion | 47 sec. | 26 sec. |
| Total CPU time to Xst completion | 46.73 sec | 25.94 sec. |
| Multipliers | 12 | 48 |
| Adders/Subtractors | 43 | 96 |
| Registers | 56 | 96 |
| Multiplexers | 317 | 384 |
| Xors | 6 | 24 |
| Total REAL time to PAR completion | 1 mins 50 secs | 2 mins 2 secs |
| Total CPU time to PAR completion | 1 mins 34 secs | 1 mins 58 secs |

## 3. CONCLUSION

In this Paper, a novel 8 point FFT architecture using pipeline and parallel processing principle described. Finally it's compared with famous Cooley-Tukey algorithm architecture. The proposed architecture has advantage of area and resource utilization and architecture can be integrated with components which are used for OFDM application in wireless communication.